# Viscous Mechano-Electric Response of Ferroelectric Nematic Liquid

*Peter Medle Rupnik,\* Luka Cmok,\* Nerea Sebastián,\* and Alenka Mertelj*

Direct viscous mechano-electric response is demonstrated for a room-temperature ferroelectric nematic liquid, which combines large spontaneous electric polarization with 3D fluidity. The mechano-electric transduction is observed in the frequency range 1 −200 Hz via a simple demonstrator device. The liquid is placed into a deformable container with electrodes and the electric current induced by both periodic and irregular actuation of the container is examined. The experiments reveal a rich interplay of several distinct viscous mechano-electric phenomena, where both shape deformations and material flow cause changes in the electric polarization structure of a ferroelectric nematic liquid. The results show that the mechano-electric features of the material promise a considerable applicative perspective spanning from sensitive tactile sensors to energy harvesting devices.

## 1. Introduction

Ferroelectricity, albite traditionally associated with solid-state materials, is in fact a phenomenon often observed in soft matter. Long-known examples are the several cases of ferroelectric polymers,[1] elastomers[2] and ferroelectric fluids forming liquid crystals (LC) phases, with columnar,[3] smectic,[4,5] or lyotropic nematic[6] order. Only the nematic LCs are true 3D liquids that retain full mobility of the constituents, however, for lyotropic nematics the value of spontaneous electric polarization $P$ is small ($\approx 2$ nC cm$^{-2}$). Recently, a novel thermotropic ferroelectric nematic liquid state[7–14] was reported with measured electric polarization values between 2 and 7 μC cm$^{-2}$,[8,13,15–17] reaching values comparable to those commonly found in solid-state materials (10 −100 μC cm$^{-2}$).[18] In such ferroelectric nematic phase ($N_F$), not only molecules are oriented on average along a common direction, but also, their dipole moments are oriented toward the same direction resulting in spontaneous electric polarization. That is, the orientational preferred direction described by the director field[19] *n*(*r*) is coupled with the electric polarization field *P*(*r*) (volume density of molecular dipole moments), thus breaking the inversion symmetry of the nematic (N) phase.

In the $N_F$ phase, spontaneous polarization order is long-range, while the material retains the fluidity which allows it to adapt to any external geometric conditions such as thin layers or microchannels.[20,21] Additionally, the material can be easily manipulated to form polar filaments[22,23] or droplets of different shapes and sizes. The latter has been shown to exhibit complex dynamics when exposed to electric fields[24–29] and temperature gradients.[30]

This unique combination of processability and high electric polarization carries a high applicative potential. The fluid character allows for changes in the soft electric polarization structure to be generated by material flow, induced as a viscous response to applied external stress, resulting in exploitable mechano-electric phenomena. That is, time-dependent changes in the liquid shape or flow, and with them, time-dependent changes in polarization orientation should result in electrical signals. This is fundamentally different from the classic piezoelectric effect, in which the electric polarisation depends on either strain or stress exerted onto the material[31] (both in the case of solids or smectic LCs[32–35]). Hence a new distinct class of non-piezoelectric, "viscous" mechano-electric effects must be considered. It is important to note, however, that very recently also converse piezoelectric effect, i.e., electric field-induced mechanic actuation, was reported for a ferroelectric nematic liquid crystal.[36]

The "viscous" mechano-electric effects, not only constitute highly interesting new fundamental physical phenomena entailing coupling effects between flow and orientational and polar order, but are also technologically relevant. The widespread use of portable and wearable devices, particularly in the quest for healthcare applications,[37] has stimulated recent ample research on piezoelectric polymers.[38,39] Contrary to hard piezoelectric ceramics, which are typically brittle, the mechanical

P. Medle Rupnik, L. Cmok, N. Sebastián, A. Mertelj
J. Stefan Institute
Jamova cesta 39, Ljubljana 1000, Slovenia
E-mail: peter.medle.rupnik@ijs.si; luka.cmok@ijs.si;
nerea.sebastian@ijs.si
P. Medle Rupnik
Faculty of Mathematics and Physics
University of Ljubljana
Jadranska ulica 19, Ljubljana 1000, Slovenia

flexibility, easiness of processing and biocompatibility make soft polymers advantageous for numerous applications.[40–45] However, they do not exhibit such a high mechano-electric conversion rate. With high spontaneous and remanent polarization values, the ferroelectric nematic LCs are expected to sustain considerable mechano-electrical transduction while retaining maximal possible mechanical compliance.

In this contribution, we demonstrate viscous mechano-electric phenomena in the $N_F$ phase of a room-temperature ferroelectric nematic LC. For that purpose, a deformable liquid crystal cell with in-plane electrodes connected to a pick-up circuit was designed and filled with a ferroelectric liquid. Such liquids have been shown to have a soft ferroelectric character in analogy to soft ferromagnets such as mu metals.[20] The material adopted a polydomain disordered polarization structure. Mechanical actuation of the cell caused structural reorientation leading to a partially poled ferroelectric liquid layer, which resulted in clear and reproducible electric current in the circuit. Such mechano-electric transduction is analyzed in parallel to optical observations, investigating material flow and reorientation as source mechanisms. Finally, we demonstrate that the dynamic range of the response enables a broad range of effective mechano-electric conversion also from irregular manual actuation.

## 2. Experimental Section

### 2.1. Material

Liquid crystalline material used in this study (FNLC-1571 provided by Merck Electronics KGaA) exhibits three distinct thermotropic LC phases, i.e., apolar nematic phase (62 −88 °C), intermediate nematic phase $N_S$ (48 −62 °C) and ferroelectric nematic phase $N_F$ (8 −48 °C). All the experiments were performed in the $N_F$ phase at room temperature. The amplitude of $P$ was measured to be 6.0 μC cm$^{-2}$ (Section 2.3), which was comparable to the values reported for other LCs with room temperature $N_F$ phase.[22,46,47]

### 2.2. Mechano-Electric Measurements

The material was filled into a custom-made wedge LC cell composed of two rectangular glass plates coated with indium-tin-oxide (ITO) electrodes on the inner surfaces (**Figure** 1c). No additional alignment layer was applied. The bottom plate contained two electrodes (0.24 and 0.48 cm$^2$) separated by a narrow 250 μm slit (in-plane electrode placement geometry).

The two plates were fixed only on one of the short ends; here the distance between the plates was determined by a 25 μm spacer. The top plate was then actuated at the free end either manually or by a linear voice-coil actuator (Thorlabs VC063) powered by a function generator and current amplifier (Sigilent SPA1010). Force was exerted onto the top plate in order to alter the geometry and volume of the cell, causing the confined material to flow (Figure 1c). In the case, when no force was applied, i.e., initial state, the LC cell was wedge-shaped with thickness increasing along the cell by a slope of ≈1 μm mm$^{-1}$. The cell thickness above the electrode slit was ≈40 μm. The cell was only partially filled. With our cell geometry, the equilibrium surface tension forces kept the material away from the thin edge of the cell and so the material motion was induced without spillage or compression of the fluid. The capacitance of the device was measured with an LCR meter (Keysight E4980A) to be $C = 120$ nF at probing signal frequency $f = 20$ Hz. The sample was placed in a grounded metallic box with openings required for optical examination of the sample (Figure 1d). Polarising optical microscopy (POM) with crossed polarisers was employed during all the experiments and white light illumination was used. Recordings were made using FLIR color camera BFS-U3-31S4C-C or IDS grayscale camera UI-3370CP-M-GL.

The sample was subjected to periodic deformations at a wide frequency range ($\nu = 1 −221$ Hz). The amplitude of the actuation signal was adjusted so that the top plate was kept out of contact with the bottom plate. A square wave signal was used to drive the voice-coil actuator, so the actuation cycle could be divided into two stages. In the first stage, the press stage, the pressure was exerted onto the top plate, causing it to elastically deform toward the bottom plate. In the second stage, the release stage, the pressure was released and the internal stress forced the plate toward the initial state.

The in-plane electrodes were connected in parallel by a resistor (18 −327 kΩ) and the oscilloscope (PicoScope 5242D). Each of the two electrodes was connected to a separate channel on the oscilloscope and both channels shared the same reference ground (Figure 1d). Zero voltage was then set as voltage where the two signals cross each other. The top electrode was not connected to the circuit. One additional set of measurements was performed on an open circuit ($R = \infty$). Because the estimated internal resistance of the sample was comparable to or higher than the oscilloscope resistance, a voltage follower was used as shown in Figure 1e to ensure that the electric current is pumped through the sample.

Experiments with manual actuation were performed on several samples. In all the cases mechano-electric response was reproduced with similar characteristics. The experiments of frequency dependency and load resistance dependency were repeated after three months and the results were consistent.

### 2.3. Measurement of Electric Polarization

For measuring the amplitude of electric polarization $P$ the triangular-wave method[48] was employed with 20 V probing signal amplitude (110 −170 Hz). The material was filled in a 5 μm thick LC cell with ITO electrodes (flat capacitor) and no surface alignment layer. The cell was connected in series with a 470 Ω resistor. The probing signal caused a complete reorientation of $n$ in a direction perpendicular to the cell plane as indicated by the complete darkening of POM images in the case of crossed polarizers. The measured electric current was integrated above the interpolated baseline as shown in Figure 1b. The amplitude of electric polarisation was estimated with $P = e/2S$, where $S$ is the electrode area.

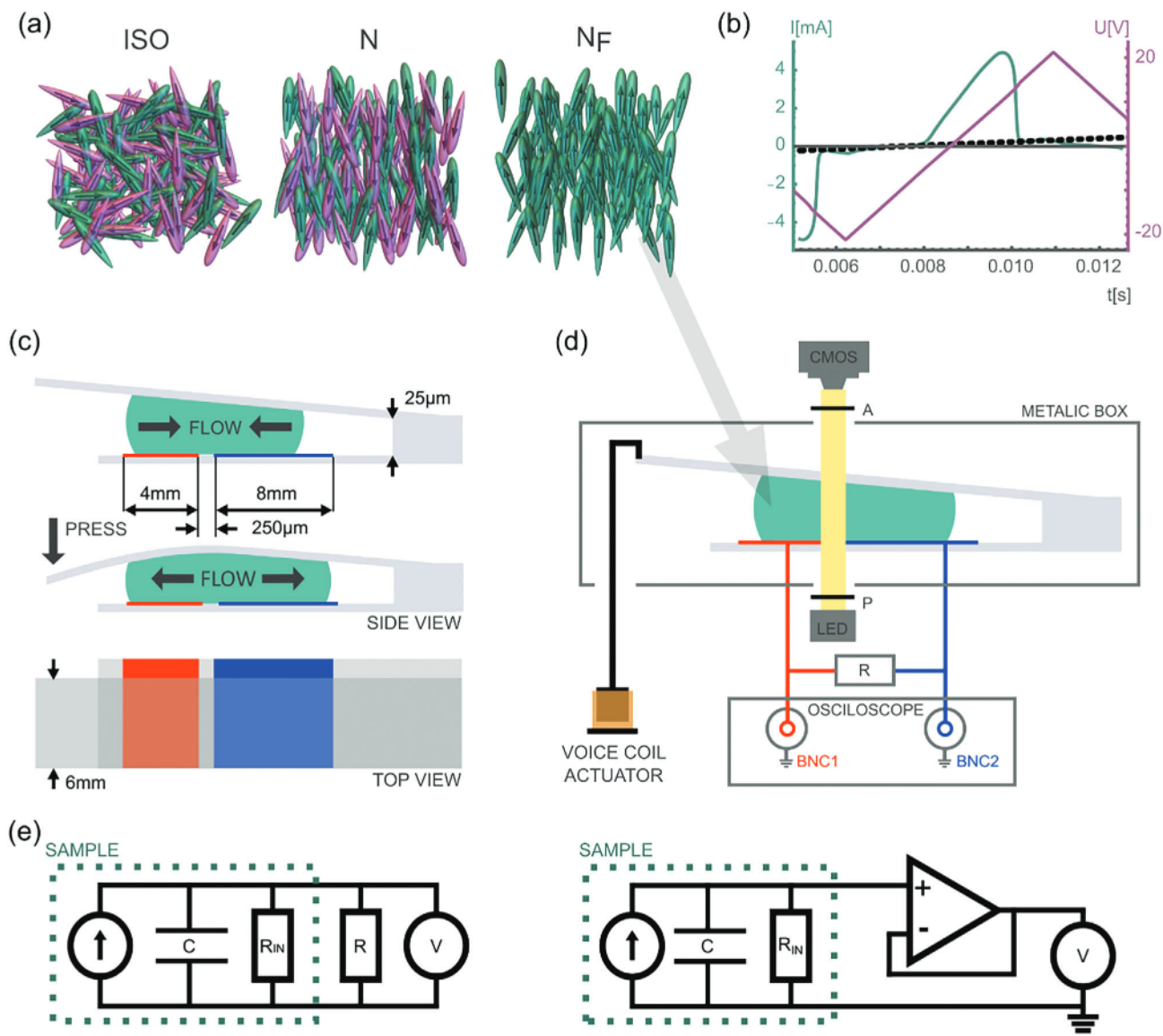

**Figure 1.** a) Schematics of isotropic phase without orientational ordering of molecules (ISO), apolar nematic phase (N), and ferroelectric nematic phase ($N_F$). Molecules and their dipoles are represented by elongated oval bodies and arrows respectively; the two colors mark the two opposite directions of dipoles. b) Example of data used for estimation of amplitude of electric polarization of ferroelectric LC. The plot shows the electric current, probing signal, and choice of a baseline in the case of 110 Hz probing frequency. c) Scheme of ferroelectric liquid crystal (green) in wedge LC cell. The first and second schemes show the side view for the initial and compressed state of the sample respectively. The third scheme shows the top view. The two electrodes are marked with red and blue colors. d) The scheme of the experimental setup shows the actuation system, the circuit, and the POM setup (LED light source polarizer P, analyzer A, and camera). e) Equivalent circuits for experiments with (left) and without (right) connected load.

## 3. Results

### 3.1. Optical Examination

Optical examination of the sample subjected to periodic mechanical stress revealed two regimes depending on the frequency and amplitude of the actuation. The first regime, with the simplest behavior, is shown in **Figure 2** for the case of 1 Hz actuation. In the initial state, the material adopted a polydomain configuration in which several complex topological defect structures were present. In Figure 2a, changes in the structure during the different times of the actuation are shown as seen by POM. Abrupt (order of 10 ms) reorientations of $n(r)$ were observed at the beginning of both stages of the actuation cycle (press and release), as indicated by the intense changes in the transmitted light intensity (Figure 2b). The reorientations then continued during the rest of the two stages but with much slower dynamics. After the initial 100 ms only slight changes in the transmitted intensity can be observed between the successive POM images in Figure 2a. During the actuation visible in-plane material flow was also observed in one or the opposite direction depending on the actuation stage as shown in Video S1 (Supporting Information). Interestingly, fast director field reorientations were often accompanied by the formation of a quasi-periodic pattern (Figure 2). These structures usually occurred soon after the inversion of the actuation stage and typically relaxed at different time scales to more homogeneous configurations, if not prevented by another inversion of the stage.

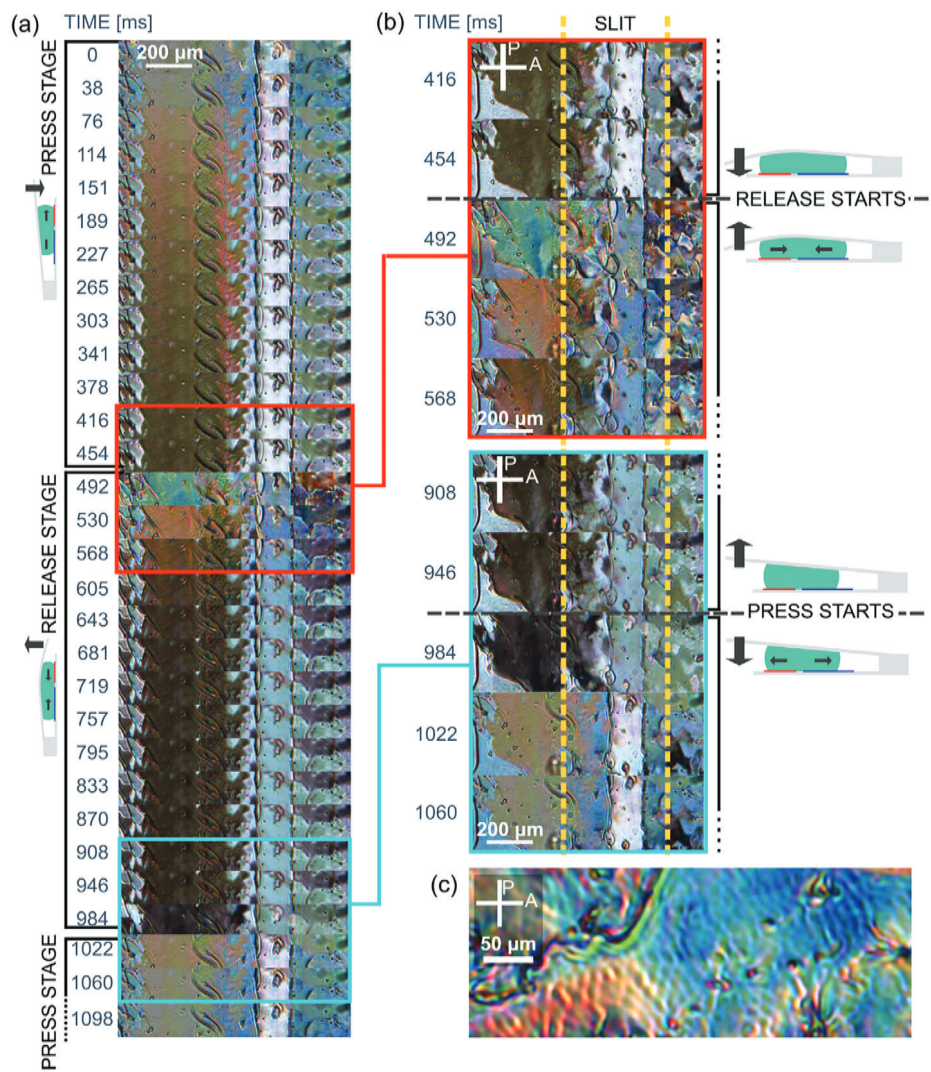

**Figure 2.** POM images of ferroelectric nematic LC in wedge LC cell during the actuation (white light source). The time interval between the two successive images is 38 ms. a) Images show the sample in the region of the slit between the electrodes at different times during a 1 Hz actuation. b) POM images showing a larger area of the sample just before and after the switch of the actuation stage. Slit location is denoted by the two yellow dashed lines. Both brightening and darkening of the image are observed depending on the domain. c) Detail from the POM images taken during the actuation showing the formation of quasi-periodic patterns.

The second regime, usually occurring at higher frequencies and higher actuation amplitudes revealed much more complex morphology as shown in Figure S1 (Supporting Information). Here the quasi-periodic pattern did not have time to relax before the next press/release stage, resulting in the generation of defects. The successive repetitions caused the material to evolve into a complex network of entangled defect lines, that relaxed back to the initial configuration state only after stopping the actuation process.

### 3.2. Different Regimes of the Electric Response

The experimental setup, as described in Section 2.2, can be represented by an equivalent circuit depicted in Figure 1e. The sample is represented by a current generator, capacitor, and resistor connected in parallel. The generated electric current corresponds to the time derivative of the effective charge, due to material movement and electric polarisation reorientations $I_P(t) = de_P/dt$. The resistor represents the internal resistance $R_{IN} \gg R$, which is much higher than the load resistance and can be thus set to infinity. Finally, with nonuniform director configurations and irregular shape deformations, the time evolution of sample capacitance could not be addressed, therefore, for simplicity, we approximate the capacitance with a constant $C$. Consequently, this highly reduced model does not account for the effects of dielectric anisotropy of the material and cannot be expected to completely reproduce all the dynamic features of the mechano-electric response. With these considerations, the circuit is described with the following differential equation:

$$0 = I_P + C\frac{dU}{dt} + \frac{U}{R} \quad (1)$$

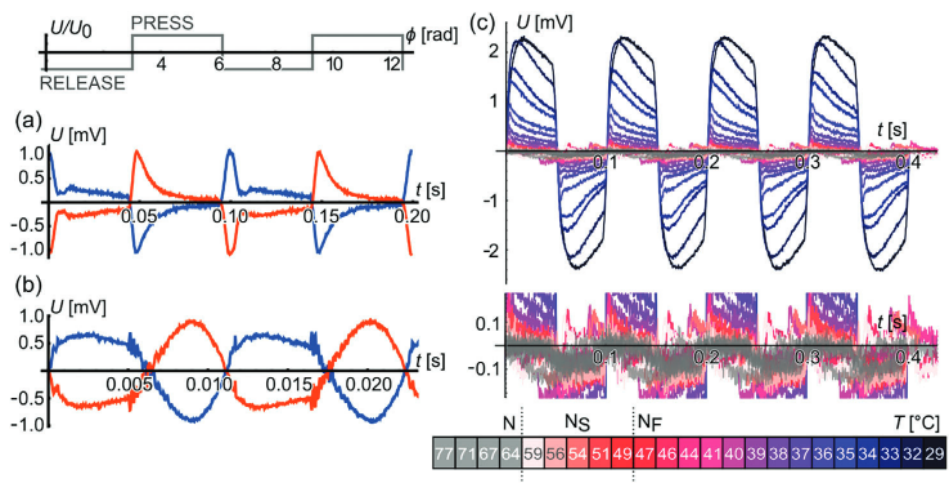

**Figure 3.** Two examples of electric response to mechanical action at room temperature at two different actuation frequencies a) 10 Hz and b) 90 Hz. The top grey plots show the normalized voltage applied to the voice-coil actuator and show the stage of the actuation in correspondence to the plots below. The blue and red colors distinguish between the signals from each of the electrodes. The time domain here is normalized by actuation period time ($\phi = 2\pi t \nu$). c) Temperature dependence of the mechano-electric response (voltage difference between the two signals) of a sample during cooling from 80 °C to room temperature. Grey color denotes cases without observed mechano-electric response, red colors denote electric responses with some features (short peaks) that followed the actuation events, and blue colors denote the clear mechano-electric response where the signal from one electrode closely resembled the inverted image of the signal on the other electrode. This mechano-electric response evolved gradually in the $N_F$ phase which is denoted by different shades of magenta color.

The equation has the solution for $U(0) = 0$:

$$U(t) = -e^{-\frac{t}{\tau}} \frac{1}{C} \int_0^t e^{\frac{t'}{\tau}} I_P(t') \, dt' \quad (2)$$

where $\tau = RC$ is the relaxation time of the circuit. In our simple case, $U(t)$ is the voltage difference on the electrodes $\frac{e_P(t)-e(t)}{C}$ ($e(t)$ are free charges on the electrode) and also the measured voltage drop on the load. Its behavior is highly dependent on the ratio between the characteristic time $t_P$ of the current generating process $I_P(t)$ and that of the circuit as shown in Figure S2 (Supporting Information). In this sense, we can distinguish between two limiting regimes $t_P \ll \tau$ and $t_P \gg \tau$. In the case of the former, the relaxations in measured $U(t)$ show the dynamics of the circuit, whereas the latter, the instantaneous regime, shows the dynamics of mechano-electric processes within the sample. Electric current through the load and the electric power are calculated by $I_R = U/R$ and $P_R = UI_R$, respectively.

### 3.3. Characteristics of the Mechano-Electric response

The described mechanical actuation of the sample was consistently accompanied by a well-pronounced electric signal. **Figure 3a** shows two examples of the electric potentials on the two electrodes connected with a 50 kΩ resistor for 10 and 90 Hz actuation frequencies. The frequency of the signal always corresponded to the actuation frequency and the sign of the voltage difference changed with the actuation stage (press/release), i.e., as the material flow was reversed. Each stage of the cycle was also associated with a very well-pronounced peak in electric signal followed by a slow relaxation. The derivatives of the two signals always had opposite signs and also, one signal closely resembled an inverted image of the other one. Those characteristics of the electric signals were qualitatively different from the case of the control experiment with an empty cell (Figure S3a, Supporting Information); the derivatives of the two signals had the same signs and different amplitudes. Here a simple electrostatic mechanism was used to generate much smaller electric currents (below 5 nA in peaks).

Further, we examined the mechano-electric response as a function of temperature. The sample, heated above 62 °C into the apolar nematic phase, produced no mechano-electric response (Figure 3c; Figure S3b, Supporting Information). On cooling, a small electric signal that followed the actuation frequency started to appear in the intermediate $N_S$ phase, however, the well-pronounced and consistent response with all the above-described features was established only deep in the ferroelectric nematic phase.

The shape of the signal was in general not the same for the two actuation cycle stages, which reflects the difference in the torques driving the deformations of the container (see Section 2.2). The press stage usually produced monotonous decreasing signals, while the release stage often produced more complex signals with non-monotonic features. The total electric charge obtained as integral of the electric current through the resistor was however consistent and similar for the two stages. A small parasitic electric current was measured independently from the actuation, however, this was negligible for higher frequencies (Figure S4c, Supporting Information).

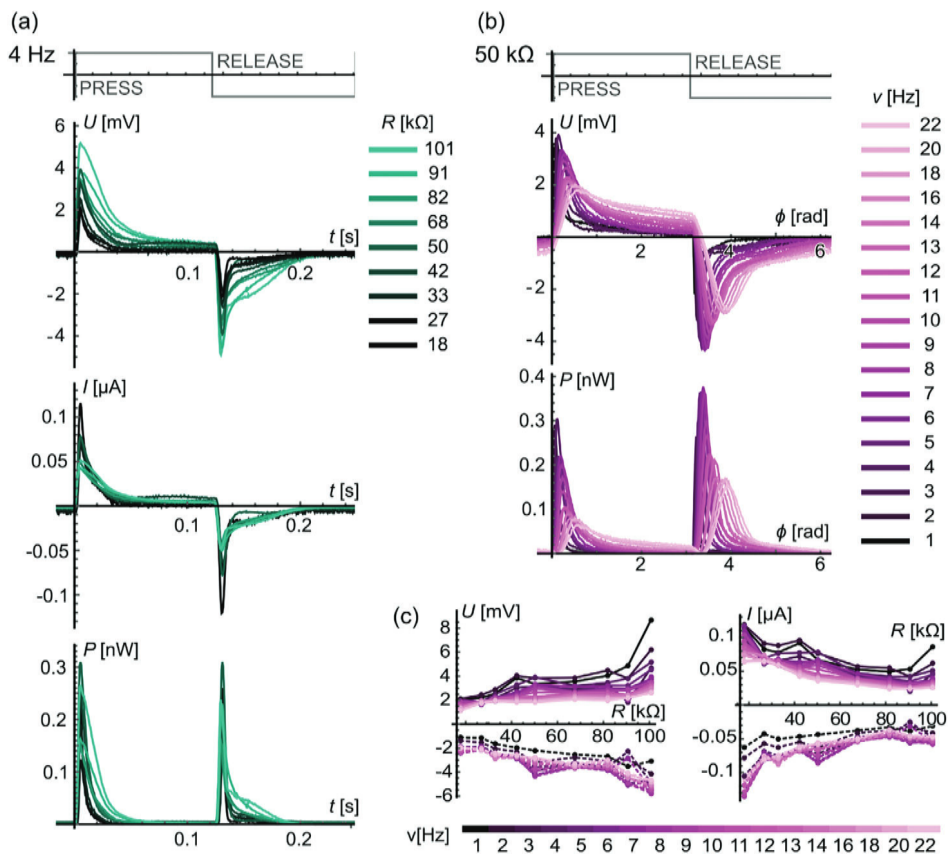

**Figure 4.** a) Time dependence of mechano-electric response at 4 Hz for different load resistances (18 −101 kΩ). b) Mechano-electric response at 50 kΩ load resistance depending on normalized time for different actuation frequencies (1 −22 Hz). The time was normalized by actuation period time ($\phi = 2\pi t \nu$). c) Load dependency of peaks of mechano-electric signals for different actuation frequencies. The maxima (press stage) peak values are connected by full and minima (release stage) by dashed lines.

The load dependence of a mechano-electric response of the system is shown in **Figure** 4a for a 4 Hz actuation frequency. Here, a voltage drop $U$ was generally increased with load resistance, and the voltage peaks were increased and widened; values of electric current $I_R$ on the other hand, were reduced with increasing load resistance (peak values are shown in Figure 4c). Qualitatively these characteristics match the electric response as calculated for the proposed simplified equivalent circuit in Section 3.2 (Figure S2, Supporting Information).

In a similar manner, the frequency dependence is shown in Figure 4b, where voltage drop $U$ and electric power $P_R$ dissipated on the 50 kΩ load resistor are shown. The shape of the response can be observed at different time scales. For low frequencies, the signal relaxation was entirely completed before the end of a press/release actuation stage (the voltage drop came close to 0). With increased frequencies, however, the signal relaxation was slower than the actuation and was interrupted by the reversal of the actuation stage. Consequently, the signal in the press stage exhibited a decreasing value of peak $U$, $I_R$, and $P_R$ (shown in Figure S4a, Supporting Information, for several load resistances). This was not the same in the more complex release stage where the maximal peak values were most often observed in the frequency range 5 −10 Hz.

The average current $\overline{I_R}$ and average power $\overline{P_R}$ spent on a chosen load as functions of actuation frequency are shown in **Figure** 5 for different load resistances. For all the cases both $\overline{I_R}$ and $\overline{P_R}$ increased with frequency in the low-frequency regime, while for high frequencies above 90 Hz, the decreasing trend was consistently observed. Here the actuation became faster than the sample response, i.e., the initial polarization reorientation. In the intermediate regime, the results were not consistent; either a single peak was observed in the frequency range of 50 −100 Hz, or a more complex behavior with several peaks was observed. The average current $\overline{I}$ increased with a decreasing load.

Additionally, we also measured the electric response of the actuated sample in an open circuit as described in Section 2.2. Relaxation time was increased significantly, however, it was finite due to the finite internal resistance of our sample. The starting dynamics of the response exhibited characteristic oscillations (Figure S4d, Supporting Information), unlike in any previous experiment.

### 3.4. Dynamics of the Electric Response

The mechanoelectric response shows relaxations on two different timescales corresponding to two distinct current-generating processes. The fast process is exponential while the slow one is roughly linear on the timescale of the measurements (Figure S5, Supporting Information). We fitted the monotone mechano-electric signals, i.e., in the low-frequency regime (1 −22 Hz) for the press stage of the actuation cycle, with the following model function:

$$f(t) = ae^{-t/t_0} + kt + n \qquad (3)$$

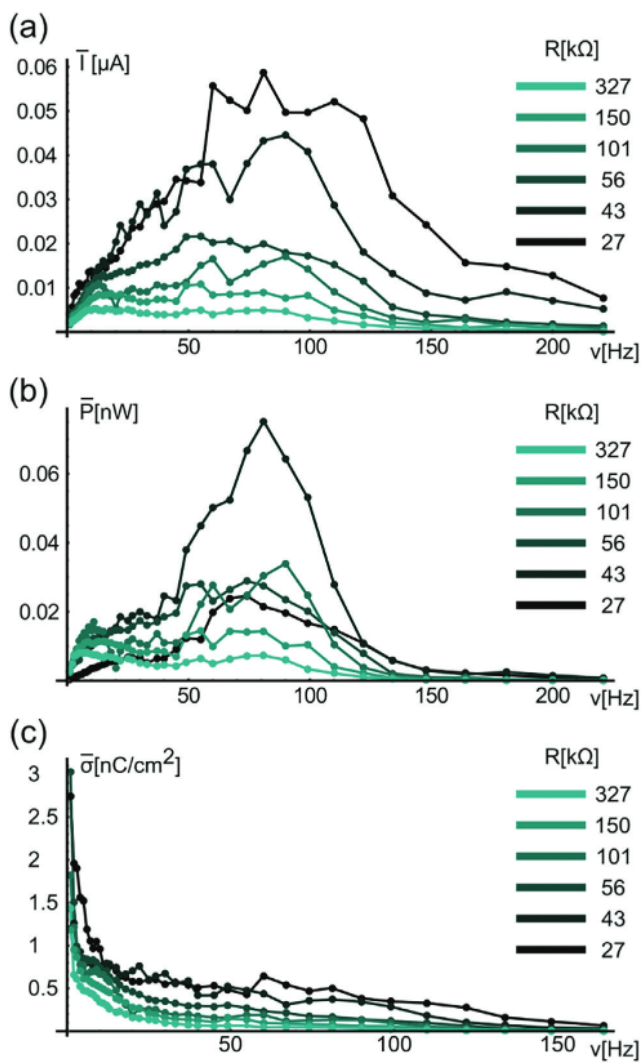

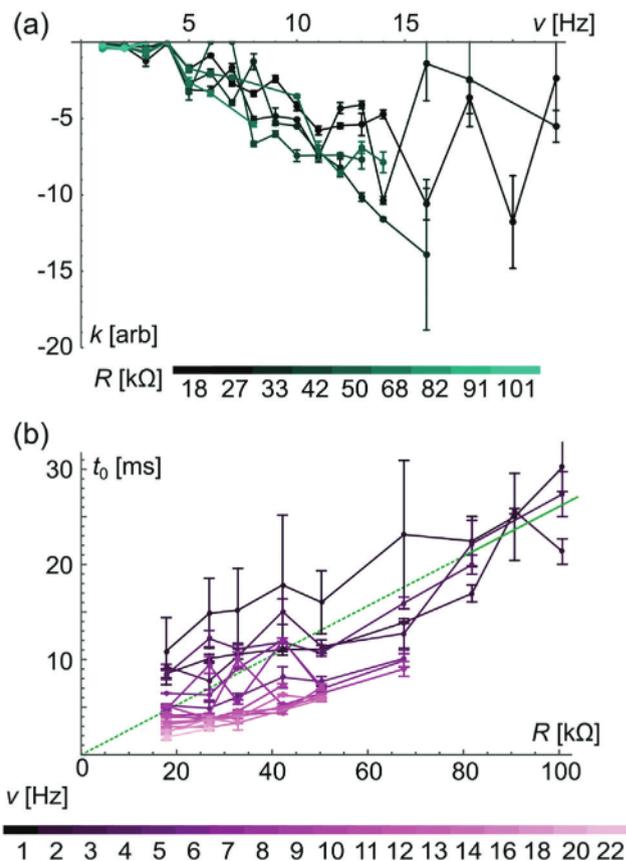

**Figure 5.** The average electric current $\overline{I_R}$ a), average electric power $\overline{P_R}$ b), and average electric charge per electrode area per actuation cycle stage $\overline{\sigma}$ c) pumped trough load as functions of actuation frequency shown for several load resistances (27 −327 kΩ). The measured values are connected by lines.

**Figure 6.** The dependence of the fit parameters $k$ and $t_0$ in the press stage of the actuation cycle. a) actuation frequency dependence of parameter $k$ for different load resistances and b) dependence of decay time $t_0$ on the load resistance $R$ for different actuation frequencies $v$. The green line shows the linear fit ($t_0 = CR$) of $t_0$ at higher load resistance $t_0(R > 80\ k\Omega)$. In (a,b) the measured values are for clarity connected by lines.

Results in **Figure 6** show two examples of how the two fitted parameters that carry information on the dynamic features of the response ($k$ and $t_0$) depend on the parameter space. In the case of the frequency dependence of the linear factor $k(v)$ Figure 6a shows roughly linear frequency dependence up to 20 Hz for several choices of load resistance. Although, due to the simplification of the model, quantitative interpretation for this relaxation is not possible, the dependency $k(v)$ points to the correlation of this dynamic parameter to the velocity of actuation (determined by the actuation frequency). This indicates that this relaxation is much slower than the circuit relaxation time ($t_{slow} \approx v^{-1} \approx 0.5 - 1\ s \gg \tau$) and therefore, shows directly the processes in the sample.

The dependence of decay time on the load resistance $t_0(R)$ is shown in Figure 6b. For all the chosen actuation frequencies $t_0$ increases with $R$. The measured values of $t_0$ are here comparable to the estimated value of circuit relaxation ($\tau \approx 2 - 12$ ms for measured $C$ and chosen load resistances $R$) and thus in principle combine both the dynamics of the faster mechano-electric process and the circuit relaxation. With small resistances, the circuit relaxation becomes faster than the current generating process $\tau < t_{fast}$ and the fitted values $t_0$ saturate toward $t_{fast}$ (instantaneous regime). From the measurements with the lowest load resistances in Figure 6b we can estimate that the current generating process fitted with the exponential decay occurs at a time scale of 1 −10 ms depending on the actuation frequency. This timescale corresponds to fast initial polarisation reorientation as optically observed in Figure 2.

In the case of higher resistances ($R > 80\ k\Omega$) in Figure 6b, however, the dependence $t_0(R)$ is approaching a linear regime suggesting that $t_0$ here is close to the circuit relaxation time $\tau$. We fitted these values with linear fit, which gave us an alternative estimation of the sample capacitance, $C = dR/d\tau \approx dR/dt_0 = 260$ nF $\pm$ 30 nF.

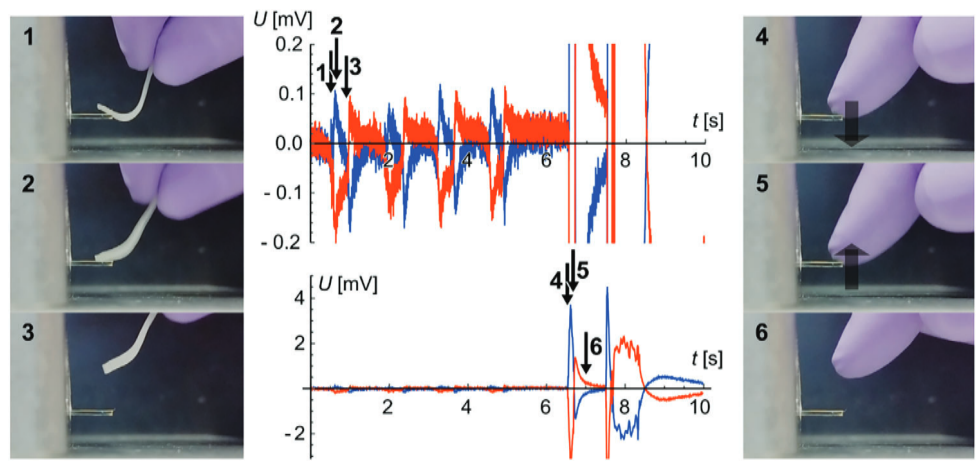

**Figure 7.** Mechano-electric response in case of irregular manual actuation. The blue and red colors distinguish between the signals from each of the electrodes. The sample is placed within a metallic box, and only the free end of the top plate comes outside the box. The left images (steps 1–3 of actuation) show the case when pressure was exerted by using a thin slab of PDMS and the right images (steps 4–6 of actuation) show the case when pressure was exerted by a fingertip. In the graphs, the mechano-electric response during the actuation process is shown in two scales to distinguish the dynamic range of the response. The arrows correspond to the time of the snapshots for the different actuation steps (1–6). The entire sequence of the response is shown in Video S2 (Supporting Information).

### 3.5. Response to Manual Actuation

The described results are very promising for the use of the $N_F$ phase in advanced applications, such as fingertip sensors or energy harvesting, where irregular activation can be expected. To test the behavior under such kind of actuation, we conducted additional experiments. In **Figure 7** (snapshots) and Video S2 (Supporting Information) (full sequence), two examples of irregular manual actuation are shown. To additionally explore the dynamic range of the response forces in two regimes were exerted on the free end of the sample's top plate. First, a thin slab of polydimethylsiloxane (PDMS) was used to tune the forces into a range between 1 and 10 mN (as estimated by applying a comparable force to a weighing scale). Although the movement of the plate was barely visible, this small disturbance was enough to cause a clear mechano-electric response. Then deformation was caused directly by the finger, which resulted in a much more significant response with complex features in the signal relaxation. The results show that already our simple device can distinguish between forces of different magnitudes and duration.

## 4. Discussion

### 4.1. Mechanism of Mechanio-Electric Transduction

The overall results show the existence of well-pronounced mechano-electric transduction specific for the $N_F$ phase that is present in a wide frequency range of mechanical actuation. Despite the complex experimental geometry and irregular material flow, the measurements of electric response were consistent and exhibited considerable qualitative matching with the calculated response of a simplified equivalent current. As interpreted, the dynamics of the measured electric signals also allowed for a reasonable comparison with morphological observations, in particular with the reorientation of the director structure.

The electric field around a droplet of a ferroelectric nematic liquid $\mathbf{E}(\mathbf{r})$ is generated mostly by electric polarisation bound charges within the material and on its surface. The first contribution is expressed as volume charge density $\rho = -\nabla \cdot \mathbf{P}$ and the second as surface charge density that corresponds to the component of electric polarisation normal to the material surface $\sigma = P_\perp$. Any movement, rearrangement, or creation/annihilation of bound electric charges causes changes in the generated $\mathbf{E}(\mathbf{r})$ and thus also the average electric potential across an electrode, placed in the proximity of the sample. In our case, because of the ferroelectric softness, the initial polydomain configuration is such that the electric field outside the material is very close to zero. Actuation causes partial poling of the liquid layer and results in voltage difference between the electrodes. Since the experimental design prevented the compression of the confined material, we deduce that no classic piezoelectric effect contributed to the electric signal. Here instead we consider the following three possible mechano-electric mechanisms.

First, already a simple rotational and/or translational motion of a poled ferroelectric droplet, without any shape or electric polarisation deformation, would cause changes in electric potential at the fixed electrodes. However, there must be an underlying mechanism which poles the soft ferroelectric material.

Second, spatial reconfiguration of $\mathbf{P}(\mathbf{r})$ is caused by any change in material droplet shape. This reconfiguration causes changes in generated $\mathbf{E}(\mathbf{r})$ (inside and outside the material). The altered depolarization field within the material causes an electric torque acting on $\mathbf{P}(\mathbf{r})$ and in the case of soft ferroelectric material, the system responds with reorientations of **P(r)** to compensate for the increase of electrostatic self-energy $F_{el} = -\frac{1}{2} \int_V \mathbf{E(r)} \, \mathbf{P(r)} \, dV$. The changes of the generated **E(r)** outside the material can be used for mechano-electric transduction. In general, this is a complex mechanism where shape dynamics couples with reorientation dynamics of **P(r)**.

Finally, any change in the droplet's shape (without compression) is necessarily accompanied by a material shear flow, which can be described by a velocity field **v(r)**.[19] A great variety of nematodynamic phenomena have been observed experimentally in the case of the apolar nematic phase from flow alignment[49] to hydrodynamic instabilities.[19] These effects arise from the coupling between the nematic orientation **n**, i.e., the average orientation of the molecular long axis, and shear flow. The physical origin of this is the coupling between translational and rotational motion of the molecules. In the $N_F$ phase, **P(r)** is parallel to **n**, so shear flow is coupled to **P(r)** which results in similar phenomena. However, this coupling is not sensitive to the sign of **P**, so the inversion symmetry breaking must come from another source, for example from the wedge shape of the layer.

The first two mechanisms are static in the sense that the polarization configuration **P(r)** depends on the sample shape and changes with its displacement. The nature of the third mechanism is dynamic in the sense that **P(r)** depends on velocity field gradients (shear flow), and changes with their temporal variations.

We expect that all the above mechanisms are present in our case. With the employed experimental geometry, however, their relative contribution to the mechano-electric signal could not be resolved. By combining optical observations and dynamical analysis at least two sets of processes occurring at two different time scales could be distinguished. The part of the signal fitted with exponential decay corresponded to the fast processes (reorientation of **n(r)** and **P(r)**) that occurred at the beginning of the actuation stages right after the reversal of flow orientation. The part of the signal fitted with the linear function corresponded to slower processes. Here the fitted parameter $k$ decreased with actuation frequency and thus the mechano-electric processes occurred in a time scale of an actuation period. In both cases, the signal was accompanied by material flow, which reaffirms the viscous nature of the contributing mechano-electric phenomena.

A very interesting study[36] was published during the preparation of this manuscript reporting on a converse mechano-electric effect in a ferroelectric nematic liquid crystal. A periodic electric signal was used to induce shape actuation of the material placed in a thin layer. Purely linear electromechanical signals were obtained in the case of high probing frequencies (above 1 kHz), thus revealing the piezoelectric character of the material. Interestingly, authors also note phenomena that could be associated with viscous response, i.e., director reorientation-induced flow, at lower frequencies, approaching our studied frequency range. Brought together, results confined in this manuscript at low frequencies and those reported above are nicely complementary. We think that the character of the mechano-electric phenomena is strongly dependent on the shear rate, which relates to the viscoelastic nature of (ferroelectric) liquids. We expect that for low shear rates, viscous mechano-electric phenomena are dominant and with increasing shear rates the piezoelectric effect gradually overcomes the viscous ones, especially in the high-frequency regime.

### 4.2. Efficiency of Mechanio-Electric Transduction

To estimate the theoretical limits of mechano-eletric transduction with ferroelectric nematic LC and evaluate the capabilities of our simple demonstrator we here consider a single abstract limiting case as shown in Figure S6 (Supporting Information). Ferroelectric nematic LC is placed on top of an electrode (located on the $xy$ plane) and we only allow for homogeneous polarization configurations along the $z$ axis so no volume bound charges are present $\nabla \cdot \mathbf{P} = 0$. The only contribution here is the surface charge density $P_\perp$ which is in the case of full reorientation of polarization (180°) twice the value of electric polarization. Under conditions comparable with our experiment (e.g., 100 Hz actuation frequency and 100 kΩ load resistance) such a thin sample could in principle generate ≈ 1 mA/cm$^2$ current density and produce for example ≈ 0.1 W of power in the case of 1 cm$^2$ electrode surface. The latter value is comparable to what one can expect to produce from commercially available solar photovoltaic panels at some reasonable natural illumination conditions.[50]

Comparing this with the maximal average electric current through the load with our sample (Figure 5) we only reached slightly below 0.01% of the theoretically estimated current and ≈0.05% of the maximal generated charge per electrode area per single reorientation in the case of slowest actuation. Further study is required to optimize the devices' geometry and improve the mechano-electric transduction.

In terms of material properties, the efficiency of hypothetic devices based on viscous mechano-electric effects concerns not only the amplitude of spontaneous electric polarisation but also how fast can **P(r)** reorient, which determines the frequency working range. Here the crucial material parameters are the orientational elastic constants of nematic LC and the viscosities of the material. In our case, the optimal working range was below 100 Hz.

Finally, the question of how much mechanical work is required to drive such type of a device falls beyond the scope of this research, since we estimate that most mechanical work in our case was used for the deformation of a rigid glass container. Indeed, in principle, one only has to overcome the surface tension forces to cause shape deformation of a free-standing ferroelectric droplet. This suggests that ferroelectric LC could be used for the construction of soft deformable tactile sensors with large sensitivity, however, further study is required to assess the mechano-electric coupling factors and general capabilities of the material.

## 5. Conclusion

In summary, we demonstrated the existence of mechano-electric transduction in a room-temperature ferroelectric nematic LC. We propose three possible mechanisms, i.e., the changes in the position of the material, changes in its shape, and changes in its internal electric polarisation structure. Translations, rotations, and deformations directly follow external action in a viscous manner, whereas polarisation structure depends on the local electrostatic field and shear flow.

Ferroelectric nematic LCs are true 3D fluids in which mechanical softness is combined with ferroelectric softness. The materials' extreme mechanical softness allows for the transduction of ambient energy in the low-frequency regime of mechanical deformation and in a regime of low external stress. As shown, the irregular manual actuation of our demonstrator provided complex and specific electric signals. We consider that the diversity of mechano-electric processes that contribute to the electric signal could allow for a fine distinction between various types of mechanical stimulus imposed onto a ferroelectric nematic liquid. The material thus has a great potential for applications such as human fingertip sensor simulation or energy harvesting from slow and irregular movements of the human body (e. g. to power batteries of heart pacers[37]). The described material capabilities are highly sought after and complement the potentialities of other existing mechano-electric transductive materials.

## Supporting Information

Supporting Information is available from the Wiley Online Library or from the author.

## Acknowledgements

The ferroelectric nematic liquid crystal material used in this work was supplied by Merck Electronics KGaA. This work was financially supported by the Slovenian Research and Innovation Agency (ARIS) (grant numbers P1-0192 and J1-50004). This study has received funding from the European Union's Horizon 2020 research and innovation programme by the project MAGNELIQ under grant agreement no 899285. The results reflect only the aurhors' view and the Commission is not responsible for any use that may be made of the information it contains.

## Conflict of Interest

The authors declare no conflict of interest.

## Author Contributions

A.M. and N.S. designed and led the study. P.M.R. performed the experiments, analyzed the results, and wrote the manuscript draft. L.C. consulted with the experiments. All authors discussed the results, reviewed the manuscript, and contributed to its final version.

## Data Availability Statement

The data that support the findings of this study are available from https://doi.org/10.5281/zenodo.11120677 and from the corresponding author upon reasonable request.